\documentstyle[12pt]{article}
\begin{document}
\begin{center}
{\large\bf ON RELATIONS BETWEEN VERTEX OPERATORS, QUASICLASSICAL
OPERATORS, AND PHASE SPACE COORDINATES}
\end{center}
\bigskip
\begin{center}
{\bf Robert Carroll\\}
{\it Mathematics Department\\University of
Illinois\\Urbana, IL 61801}
\end{center}
\bigskip
\begin{center}
{\large Abstract}
\end{center}
For certain situations we give a geometrical background
for quasiclassical KP calculations based on an explicit
connection to quantum mechanics and the collapse of coherent
states to coadjoint orbits for classical operators.
\\[1cm]\noindent {\bf 1. INTRODUCTION.}
\\[.5cm]\indent Despite the fact that soliton equations such as KdV, KP, NLS,
or DS often arise physically from water wave problems
for example, the background mathematics involves tau functions,
wave functions, Lax operators, Grassmannians, etc. and has a
combinatorial nature at times with many quantum mechanical
features.  There are also many
direct connections of soliton mathematics to quantum mechanics,
conformal field theory, quantum gravity, strings, etc. and we
mention here only [1-4;8;13;18;20;25;42;46;49]
which have some direct relation to the matters we discuss.
If one looks at this background mathematics for the KP hierarchy
for example one sees the variables $x = t_1,$ and $t_n\,\,(n\ge 2)$
arising via vertex operators in the bosonization process in representation
theory a la [20], or directly as in [9] (cf. also [8;34;46]).  In terms
of the mathematics this should then be regarded as the basic meaning;
thus for example $\partial \sim a$ and $x \sim a^{\dagger}$ with
$[a,a^{\dagger}] = 1$.  We do not dwell here upon the origins of KP
and KdV equations via algebraic curves, which should also be regarded
as another fundamental point of departure (cf. [8;30;31;38]).  It
seems especially important to regard the $t_n,\,\,n \ge 1,$ in this
way since they play different roles in different theories.  For example
their role in theories of (p,q) minimal matter coupled to 2-D gravity
seems no less (and perhaps more) fundamental than their role in water
waves.  In such theories x plays the role of a cosmological constant
($\partial \sim$ a puncture operator), the $t_n$ are coupling constants
($\sim$ deformation parameters), and the tau function corresponds to
a partition function.  Thus the $x,\,\,t_n\,\,(n \ge 2)$ do have a
coordinate aspect but it is not clear why techniques of fast and slow
variable scaling and averaging (which arise naturally in water wave
problems), are appropriate here.  Also, there is a sense in which one
can think of the coupling constants as emerging out of the phase space
of $(x,\partial)$ (cf. [49]) so their coordinate nature is somewhat
secondary.  Another feature here is that in passing to quasiclassical
limits in physics or to dispersionless limits in water waves via a
scaling $\epsilon t_n = T_n\,\,(\epsilon x = X)$ the background
mathematics passes from quantum mechanical to classical (or
quasiclassical) and the averaging or scaling procedure has nothing
intrinsically quantum mechanical about it, nor anything geometrical.
Hence we were led to try to provide an underlying
phase space or geometrical context in which to view
the scaling mechanism.  In this paper we show how, in certain
situations at least, the scaling can be related to quantum mechanical
procedures in [17;47] where coherent states collapse onto a
coadjoint orbit (cf. also [48]).  This seems, to me at least,
much more satisfactory than thinking of fast and slow variables,
or averaging, which appear too mysterious as a general directive.  The coherent
state idea can be carried further in the soliton context as indicated
in [34;35] but we do not pursue this here (see also
comments and suggestions in section 4).  We also give some further
results on the semiclassical action principle of [3;4], relating
the integrand to a limit of the Sato equation of KP theory.
Finally some heuristic comments are made concerning relations of
the Maslov canonical operator to dressing ideas and its possible
use in quasiclassical or dispersionless soliton mathematics.
\\[3mm] \indent In terms of providing a theme in our presentation
we suggest starting with the boson operators $\partial \sim a,
x \sim a^{\dagger}$ acting on a vacuum $|0> = 1$.  One creates then
states via $|n> = (a^{\dagger})^n|0>/\sqrt{n!}$ for example or
coherent states via $|z> = D(z)|0>$ as in Appendix A.  One is working
in a Fock space $\hat{{\cal B}}$ here with no a priori reference to
physics.  Then one asks about classical and/or quasiclassical objects
related to $\hat{{\cal B}}$ and the state vectors above.  For this
one needs some sort of physical object having a classical counterpart
so we create such an object via $\hat{q} = (a + a^{\dagger})/\sqrt{2}$
and $i\hat{p} = (a - a^{\dagger})/\sqrt{2}$ with $\hat{Q} \sim
\sqrt{h}\hat{q}$ and $\hat{P} \sim \sqrt{h}\hat{p}$.  This leads to
the development in the text.  Thus given $\tilde{g}_h$ the Lie algebra
generated by $e_1 = i\hat{P}/h, e_2 = i\hat{Q}/h,$ and $e_3 = i/h$
with corresponding Weyl-Heisenberg group $G_h = \{U_h = exp[\frac{i}
{h}(u + \frac{1}{2}\pi\xi)]U(\pi,\xi)\}$ where $U(\pi,\xi) \sim
exp[\frac{i}{h}(\pi\hat{Q} - \xi\hat{P})] \sim D(z), \,\,z =
\frac{1}{\sqrt{2h}}(\xi + i\pi)$, one generates coherent states
$|u> = U_h|0>_h$ ($|0>_h$ being a peaked vacuum for $\hat{Q}$).
This leads to coadjoint orbits $\Gamma = \{\zeta = Ad^{*}_u\zeta_0,
\,\,u \in G_h,\,\,\zeta_0 \in \tilde{g}^{*}_h$ fixed$\}$ and for
$lim <0|\frac{h}{i}\hat{\Lambda}|0> = <\zeta_0,\lambda>$ ($\lambda
\sim (\pi,\xi,u)$ in $\tilde{g}_h; \hat{\Lambda} \sim (\pi e_1,
-\xi e_2, u e_3)$) one has $<u|\frac{h}{i}\hat{\Lambda}|u>_h \to
<\zeta_0,Ad_{u^{-1}}(\lambda)> = <Ad^{*}_u\zeta_0,\lambda>$.  The
variables $(\pi,\xi)$ provide coordinates on $\Gamma$ and there is
a symplectic structure, etc.  Such operators $\hat{\Lambda}$ generate
classical operators via $\hat{A} = \int d\lambda f(\lambda)
exp[h\hat{\Lambda}]$ and covariant symbols $A_h(u) = <u|\hat{A}|u>_h$
lead to functions $a(\zeta)$ on $\Gamma$.  Expectation values of
$\frac{h}{i}\hat{\Lambda}$ distinguish classically inequivalent states
and coherent states thus collapse to $\Gamma$ via an equivalence
relation.  Therefore $\Gamma$ provides a classical geometric underpinning
for $\hat{{\cal B}}$ via the ad hoc prescription of $(\hat{p},\hat{q})$
based on $(a,a^{\dagger})$.  This is all in the mathematics and the
procedure works for any $(a,a^{\dagger})$.  The only physical objects
are the contrived $(\hat{p},\hat{q})$.  Now once this is set up one
can refer to the results of Hepp involving the natural convergence
$\hat{Q} \sim \sqrt{h}\hat{q} \to \xi, \hat{P} \sim \sqrt{h}\hat{p}
\to \pi$, etc. to discuss quasiclassical operators defined by (4.38)
via $\hat{D}_{\epsilon}(z) = exp[\frac{i}{\epsilon}(\tilde{\pi}\hat{Q}
- \tilde{\xi}\hat{P})]\,\, (\tilde{\xi} = \sqrt{2}\xi, \tilde{\pi} =
\sqrt{2}\pi)$ where $\epsilon^2 \sim h$.  These correspond in fact
to certain vertex operator situations from dKP theory and allow one
to give a geometrical underpinning to dispersionless or quasiclassical
limits via the phase space $\Gamma$.  The original $\partial$ and $x$,
determining $a,a^{\dagger}$ in this particular situation, can then
be recovered in a scaled form $x \to X/\epsilon,\,\,\partial \to
\epsilon\partial$ via (4.39) in terms of $(\xi,\pi)$.  This makes the
scaling procedure part of a general mathematical framework related
to an underlying phase space.  Let us mention also that the idea of
classicalization is of course rather more complicated than simply
letting h tend to 0 and an extensive theory of the quantum-classical
correspondence is developed in [50].  Generally there are a number
of classical systems corresponding to a given quantum system and in
[50] for example one works with a classicalization in terms of the
removal of quantum fluctuation effects in the physical observables,
which is not necessarily the usual limit $h\to 0$. This really has
no effect on the basic mathematical theory involved here in which
quasiclassical operators are defined and related to the dispersionless
KP theory for example.  On the other hand in terms of suppression
or accomodation of quantum fluctuations ($\sim$ rapid oscillations)
the ideas of [50] may well have meaning and applicability in
discussing weak and generalized solutions of KdV type water wave
problems (cf. [27]).
\\[1cm]\noindent {\bf  2. SOME QUANTUM MECHANICS AND COHERENT STATES.}
\\[.5cm]\indent We begin with a sketch of some ideas in [17].  Consider a
classical system
\medskip
$$\leqno(2.1) \quad {\cal H} = \pi^{2}/2m + V(\xi); m\dot {\xi} = \pi;
\dot {\pi} = -V'(\xi);$$
$$\xi(\alpha,0) = \xi; \pi(\alpha,0)=\pi;
\alpha = \frac{\xi + i\pi}{\sqrt{2}}$$
\noindent The associated quantum mechanical system has the form
\medskip
$$\leqno(2.2) \quad ih\psi_{t} = -\frac{h^{2}}{2m}\psi_{qq} + V\psi;
H_{h} = \frac{p^{2}_h}{2m} + V_{h}; ip_{h} = h\partial_{q};$$
$$\psi_{t} = U_{h}\psi_{0}; V_{h} \sim{V(h^{\frac{1}{2}}\xi)};
U_{h} = exp(-it\frac{H_{h}}{h})$$
\noindent (here $H_{h}$ is some selfadjoint extension of the symmetric
operator indicated and we use h for Dirac's slash h).  In this
context $q \sim{x}, \hat {q}\sim{operator},$ and
\medskip
$$\leqno(2.3)\quad p_{h} = \sqrt{h}p; q_{h} = \sqrt{h}q;
\hat {p} = -i\partial_{X}; [\hat {q},\hat {p}] = [x,-i\partial_{X}] = i;$$
$$[\hat {q}_{h},\hat {p}_{h}] = hi; a = \frac{(\hat {q} + i\hat
{p})}{\sqrt{2}};
a^{\dagger} = \frac{(\hat {q} -i\hat {p})}{\sqrt{2}};
\hat {q} = \frac{(a + a^{\dagger})}{\sqrt{2}};
i\hat {p} = \frac{(a - a^{\dagger})}{\sqrt{2}}$$
\medskip
\noindent We note that
\medskip
$$\leqno(2.4)\quad a = a_{h} = \frac{(\hat{q}_{h} + i\hat{p}_{h})}{\sqrt{2h}};
a^{\dagger} = a^{\dagger}_{h} = \frac{(\hat{q}_{h} -i\hat{p}_{h})}{\sqrt{2h}}$$
\medskip
\noindent and this is related to scaling as follows.  Let $q\to\epsilon q =
Q, \partial_{q}\to\epsilon\partial_{Q}, q = \frac{Q}{\epsilon}$ (the situation
for quasiclassical limit calculations in soliton mathematics).  Then
\medskip
$$\leqno(2.5)\quad a = \frac{(\frac{\hat {Q}}{\epsilon}
 + \epsilon\partial_{Q})}{\sqrt{2}}
= \frac{(\hat {Q} + i\hat {P})}{\epsilon\sqrt{2}}= a(\epsilon);
i\hat {P} = \epsilon^{2}\partial_{Q}; a^{\dagger} =
\frac{(\hat {Q} - i\hat {P})}{\epsilon\sqrt{2}} = a^{\dagger}(\epsilon)$$
\medskip
\noindent where $a(\epsilon)\sim{a_{h}}$ for $\epsilon^{2} = h.$  This
leads to $\hat {Q}\sim\hat {q}_{h}$ and $\hat {P}\sim\hat{p}_{h}$ with
$[\hat {Q},\hat {P}] = i\epsilon^{2}$ and (note $\epsilon\hat{q} = \hat{Q},
\epsilon\hat{p} = -i\epsilon\partial_{q} = -i{\epsilon}^{2}\partial_{Q})$
\medskip
$$\leqno(2.6)\quad \frac{\hat {Q}}{\epsilon} =
\frac{(a(\epsilon) + a^{\dagger}(\epsilon))}{\sqrt{2}};
[a,a^{\dagger}] = [a_{h},a^{\dagger}_{h}] = 1;
\frac{i\hat{P}}{\epsilon} = \epsilon\partial_{Q} =
\frac{(a(\epsilon) - a^{\dagger}(\epsilon))}{\sqrt{2}}$$
\medskip
\indent Consider now $z = h^{-\frac{1}{2}}\alpha$ and $U(p,q)\sim D(z)$
(cf. (A.5)-(A.6)) with $U^{*} = U^{\dagger}\sim D^{\dagger}(z) =
D(-z)\sim U(-p,-q).$  Look at a typical expression $(|h^{-\frac{1}{2}}|\alpha>
=
U(h^{-\frac{1}{2}}\alpha)|0>)$
\medskip
$$\leqno(2.7)\quad \Xi = <h^{-\frac{1}{2}}\alpha|(\hat {q} -
\frac{\xi}{\sqrt{h}})
(\hat {p} - \frac{\pi}{\sqrt{h}})|h^{-\frac{1}{2}}\alpha> =$$
$$<0|U^{*}(h^{-\frac{1}{2}}\alpha)(\hat {q} - \frac{\xi}{\sqrt{h}})
UU^{*}(\hat {p} - \frac{\pi}{\sqrt{h}})U|0>$$
\medskip
\noindent Using (A.8) one has $(a = a_{h})$
$$\leqno(2.8)\quad U^{*}(z)[\frac{(\hat{q} + i\hat{p})}{\sqrt{2}}]U(z) =
\frac{(\hat{q} + i\hat{p})}{\sqrt{2}} +
\frac{(\xi + i\pi)}{\sqrt{2h}}$$
\medskip
\noindent which implies
\medskip
$$\leqno(2.9)\quad U^{*}(z)\hat {q}U(z) = \hat {q} +
\frac{\xi}{\sqrt{h}}; U^{*}(z)\hat {p}U(z) = \hat {p} +
\frac{\pi}{\sqrt{h}}$$
\medskip
\noindent Consequently from (2.7)
\medskip
$$\leqno(2.10)\quad \Xi = <0|\hat {q}\hat {p}|0>$$
\medskip
\noindent This implies
\medskip
$$\leqno(2.11)\quad <h^{-\frac{1}{2}}\alpha|(\hat {q}_{h} - \xi)
(\hat {p}_{h} - \pi)|h^{-\frac{1}{2}}\alpha>\to 0;
<h^{-\frac{1}{2}}\alpha|\hat {Q}\hat {P}|h^{-\frac{1}{2}}\alpha>
\to \xi \pi$$
\medskip
\noindent Thus the scaled variables $Q\sim\hat {q}_{h}\to \xi,$ and
$\hat {P}\sim\hat {p}_{h}\to \pi,$ where $\xi, \pi$ correspond to
classical values of $\hat {q}, \hat {p}.$
\\[3mm]\indent Now we want to relate this to [47] where it is shown how
coherent states collapse onto coadjoint orbits as $h\to 0.$  Thus,
using the notation of [47]
\medskip
$$\leqno(2.12)\quad U_{h}(p,q,u) = e^{\frac{iu}{h}}
e^{\frac{ip\hat {Q}}{h}}e^{-\frac{iq\hat{P}}{h}}$$
\medskip
\noindent with $e_{1} = \frac{i\hat {P}}{h}, e_{2} = \frac{i\hat {Q}}{h},$
and $e_{3} = \frac{i}{h}$ generating a Lie algebra $\tilde {g}_{h}$
($[e_{1},e_{2}] = e_{3}\sim [\hat {Q},\hat {P}]$ = ih).  Evidently,
via the Baker-Campbell-Hausdorff (BCH) formula for the present situation
\medskip
$$\leqno(2.13)\quad e^{A}e^{B} = e^{\frac{1}{2}[A,B]}e^{A + B} =
e^{[A,B]}e^{B}e^{A}$$
\medskip
\noindent (valid when [[A,B],A] = [[A,B],B] = 0) one has (cf. (A.6))
\medskip
$$\leqno(2.14)\quad U_{h}(p,q,u) = e^{\frac{i}{h}(u + \frac{1}{2}pq)}U(p,q)$$
\medskip
\noindent In [17] one writes $\lambda = \sum \lambda^{i}e_{i}\in\tilde{g}_{h}$
and $\xi = \sum \xi_{i}e'_{i}\in\tilde{g}^{*}$ (for Lie groups,
coadjoint orbits, etc. see e.g. [5;7;8;21;33]).  The Weyl-Heisenberg
(WH) group $G_{h}$ generated by the $U_{h}$ has Lie algebra $\tilde{g}_{h}$
and one has coadjoint orbits
$\Gamma = {Ad^{*}_{u}\zeta_{0}}$ for $\zeta_{0}\in\tilde{g}^{*}_{h}$
fixed.  Thus a coherent state $|u> = U_{h}|0>$ gives rise to a
point $\zeta = Ad^{*}_{u}\zeta_{0}\in\Gamma (|0>$ will be a peaked
vacuum as in (A.5)) and the set [u] of equivalence classes of
coherent states corresponds to $\Gamma$.  The tangent space
$T_{\zeta}(\Gamma)\subset\tilde{g}^{*}$ at $\zeta$ is now generated by
$ad^{*}_{\lambda}(\zeta), \lambda\in\tilde{g}_{h} (<ad^{*}_{\lambda}(\zeta),
\eta> = -\hfil\newline
<\zeta,[\lambda,\eta]>)$ and $T^{*}_{\zeta}(\Gamma)$
can be identified with equivalence classes $[\lambda], \lambda\in
\tilde{g}_{h},$ via $[\lambda] = \{\lambda'\in\tilde{g}_{h};
ad^{*}_{\lambda}\zeta = ad^{*}_{\lambda'}\zeta\}.$  Let $H_{\zeta}$ = isotropy
group of $\zeta = \{u\in G; Ad^{*}_{u}\zeta = \zeta\}$
so $\Gamma\sim \frac{G}{H_{\zeta}}.$
Explicitly, for the WH group coadjoint orbits correspond to planes
$\zeta_{3}$ = constant $(\not = 0)$ since
\medskip
$$\leqno(2.15)\quad Ad^{*}_{u}(\sum \zeta_{i}e'_{i}) =
\sum \hat {\zeta}_{i}e'_{i};
\hat {\zeta}_{3} = \zeta_{3};
\hat {\zeta}_{1} = \zeta_{1} + p\zeta_{3};
\hat {\zeta}_{2} = \zeta_{2} + q\zeta_{3}$$
\medskip
\noindent Relabeling $\zeta_{1}\sim p, \zeta_{2}\sim q$ one has
natural coordinates (p,q) on $\Gamma$ and a symplectic structure
with Poisson brackets, etc. can be written down.  In particular
$\{f,g\} = \zeta_{3}[f_{p}g_{q} - f_{q}g_{p}].$
\\[3mm]\indent Now associated to $\lambda$ = (p,q,u) one has the operator
$\hat {\Lambda} = (\frac{ip\hat {Q}}{h},-\frac{iq\hat {P}}{h},
\frac{iu}{h})$ and $h\hat {\Lambda}$ is a "classical operator",
i.e.,$<\zeta_{0},\lambda> = lim (\frac{1}{i})<0|h\hat {\Lambda}|0>$ exists.
Classical operators are then defined via a symbolic representation
\medskip
$$\leqno(2.16)\quad \hat{A} = \int d\lambda f(\lambda)e^{h\hat {\Lambda}}$$
\medskip
\noindent $(exp(h\Lambda)\sim U_{h}(hp,hq,hu)$ via BCH).  The covariant
symbol (following Berezin - cf. [39]) $A_{h}(u)$ is defined as the
set of coherent state expectation values
\medskip
$$\leqno(2.17)\quad A_{h}(u) = <u|\hat {A}|u>_{h}$$
\medskip
\noindent where $|u>_{h}\sim U_{h}|0>_{h}, |0>_{h}$ being the
peaked state vacuum which we simply write as $|0>$ when no confusion
can arise.  Given e.g.
\medskip
$$\leqno(2.18)\quad lim (\frac{h}{i})<0|\hat {U}^{-1}\hat {\Lambda}\hat {U}|0>
= <\zeta_{0},Ad_{u^{-1}}(\lambda)> = <Ad^{*}_{u}\zeta_{0},\lambda>$$
\medskip
\noindent we distinguish classically equivalent states as those $|u>$
mapped onto a given point $\zeta\in \Gamma$ (i.e. expectation values
of $h\hat{\Lambda}$ distinguish classically inequivalent states).  Then for
any classical operators $\hat {A}, \hat {B}$ as in (2.16) one has
\medskip
$$\leqno(2.19)\quad A_{h}(u)\to a(\zeta); (AB)_{h}(u)\to a(\zeta)b(\zeta);
\frac{i}{h}[A,B]_{h}(u)\to \{a(\zeta),b(\zeta)\}$$
\medskip
\noindent A more or less precise development of all this is given
in [47] (cf. also [16;48]).
\\[1cm]\noindent {\bf 3. SOME SOLITON VERSIONS.}
\\[.5cm]\indent In dealing with vertex operators in soliton mathematics one
will
encounter terms $exp(x\lambda - \lambda^{-1}\partial)$ which can
be viewed in several ways. As indicated in Appendix A, essentially we will
gratuitously introduce a coordinate representation and coherent states
and treat $(\partial,x)$ as boson operators $(a,a^{\dagger})$.
First scale $x\to \epsilon x = Q,
\partial_{x}\to \epsilon\partial_{Q}$ to obtain (cf. (2.5)
\medskip
$$\leqno(3.1)\quad \lambda\frac{\hat {Q}}{\epsilon} -
\lambda^{-1}\epsilon\partial_{Q} = \frac{(\lambda - \lambda^{-1})}{\sqrt{2}}
a(\epsilon) + \frac{(\lambda + \lambda^{-1})}{\sqrt{2}}a^{\dagger}(\epsilon);
a(\epsilon) = \frac{1}{\sqrt{2}}(\hat {Q} +\epsilon^{2}\partial_{Q});$$
$$a^{\dagger}(\epsilon) = \frac{1}{\epsilon\sqrt{2}}(\hat {Q} -
\epsilon^{2}\partial_{Q}); \frac{\hat {Q}}{\epsilon} =
\frac{1}{\sqrt{2}}(a(\epsilon) + a^{\dagger}(\epsilon));
\epsilon\partial_{Q} = \frac{1}{\sqrt{2}}(a(\epsilon) -
a^{\dagger}(\epsilon))$$
\medskip
\noindent One can generate peaked states exactly as before, using
$a(\epsilon)$ and $a^{\dagger}(\epsilon) (\epsilon^{2}\sim h).$  Writing
$\hat {P} = \epsilon^{2}\partial_{Q}$ (no i) we can set
\medskip
$$\leqno(3.2)\quad \frac{(\lambda\hat{Q} - \lambda^{-1}\hat{P})}{\epsilon} =
\alpha a^{\dagger}(\epsilon) + \beta a(\epsilon); \alpha =
\frac{(\lambda + \lambda^{-1})}{\sqrt{2}}; \beta =
\frac{(\lambda - \lambda^{-1})}{\sqrt{2}}$$
\medskip
\noindent In particular if $\lambda\in S^{1}$ so $\lambda^{-1} =
\bar {\lambda}$ one obtains
\medskip
$$\leqno(3.3)\quad \frac{(\lambda\hat{Q} - \lambda^{-1}\hat{P})}{\epsilon} =
\sqrt{2}(Re(\lambda)a^{\dagger}(\epsilon) + iIm(\lambda)a(\epsilon))$$
\medskip
\noindent which however we do not exploit. \hfil\newline
\indent Alternatively one can think of $x\sim a^{\dagger}$ and
$\partial\sim a$ with (cf. (2.5))
\medskip
$$\leqno(3.4)\quad a = \frac{(\hat {q} + \partial_{q})}{\sqrt{2}};
a^{\dagger} = \frac{(\hat {q} - \partial_{q})}{\sqrt{2}}; a\sim a(\epsilon) =
\frac{(\hat{Q} + \epsilon^{2}\partial_{Q})}{\epsilon\sqrt{2}};$$
$$a^{\dagger}\sim a^{\dagger}(\epsilon) = \frac{(\hat {Q}
-\epsilon^{2}\partial_{Q})}{\epsilon\sqrt{2}};
\lambda a^{\dagger} - \lambda^{-1}a = \frac{[(\lambda - \lambda^{-1})
\hat {Q} - (\lambda + \lambda^{-1})\epsilon^{2}\partial_{Q}]}
{\epsilon\sqrt{2}}$$
\medskip
\noindent For $\lambda\in S^{1}$ this becomes ($a,\,\,a^{\dagger}$ are
based on $(\partial,x))$
$$\leqno(3.5)\quad \lambda a^{\dagger} - \lambda^{-1}a =
\lambda a^{\dagger}(\epsilon) - \bar{\lambda}a(\epsilon) =
\frac{i\sqrt{2}[Im(\lambda)\hat{Q} + iRe(\lambda)\epsilon^{2}\partial_{Q}]}
{\epsilon}$$
\medskip
\noindent Looking at (A.7) one has $za^{\dagger}_{h} - \bar{z}a_{h} =
(\frac{i}{h})(p\hat{Q} - q\hat{P})$ with $z = \frac{(q + ip)}{\sqrt{2h}}$
and we think of $\epsilon^{2}\sim h$ so $\epsilon^{2}\partial_{Q}\sim
i\hat{P}.$
Then compare with (3.5) rewritten as
\medskip
$$\leqno(3.6)\quad \frac{(\lambda a^{\dagger}(\epsilon) -
\bar {\lambda}a(\epsilon))}{\epsilon\sqrt{2}} = \frac{i[Im(\lambda)\hat {Q} -
Re(\lambda)\hat {P}]}{\epsilon^{2}}$$
\vskip .2in
\noindent {\bf THEOREM 3.1.} $\;$ For $\lambda\in S^{1}$ (3.5) is quantum
mechanical in nature with $p = Im(\lambda), q = Re(\lambda) \; (p^{2} +
q^{2} = 1),$ and $z = \frac{(q + ip)}{\epsilon\sqrt{2}} =
\frac{\lambda}{\epsilon\sqrt{2}}.$  The $\hat {Q}$ operator however arises
as in (3.4) and is not directly a scaling of $\hat {x}$ (cf. (4.21) for
connections).
\vskip .2in
\noindent {\bf REMARK 3.2.} $\;$ Let now $\lambda$ be general and look at
(3.2) and (3.4), with $\epsilon^{2}\partial_{Q} = \hat {P}.$  For both
cases one will have the right scaling to
fit in the framework of (3.6) $(\frac{(\alpha,\beta)}{\epsilon}$ in
(3.2) or $\frac{\lambda}{\epsilon}$ in (3.4)) but there seems to be
no way to phrase this in terms of unitary operators if $\lambda
\not\in S^{1}$ (one would want say $\lambda$ real in (3.2) and
$(\alpha,\beta)$ real in (3.4)).
Thus we will develop the situation of Theorem 3.1.  In this case
we note that the measure $d\mu$ in (A.7) is inappropriate and one
thinks rather of $\oint\frac{dz}{z}$ suitably normalized.  For example
consider from (A.5)
\medskip
$$\leqno(3.7)\quad u_{n}(z) = <n|z> = e^{-\frac{1}{2}}\frac{z^{n}}
{\sqrt{n!}}$$
\medskip
\noindent One can then generate an orthonormal set $w_{n} = c_{n}
u_{n} = e^{\frac{1}{2}}\sqrt{n!}u_n = z^n$ so that $(z\in S^{1})$
$$\leqno(3.8)\quad <w_{n}|w_{m}> = \bar{c}_{n}c_{m}\frac{ce^{-1}}
{2i\pi}\oint\frac{\bar{z}^{n}z^{m}dz}{z\sqrt{n!}\sqrt{m!}} =
\delta_{mn}$$
\\[6mm]\noindent {\bf 4. CONNECTIONS TO DISPERSIONLESS LIMITS.}
\\[.5cm]\indent We refer here to [2-4;6;8;23;24;42;43] for background.  For
classical KP one has a Lax operator $L = \partial + \sum^\infty_{1}
u_{n+1}\partial^{-n}$ and a gauge operator $W = 1 + \sum^\infty_{1}
w_{n}\partial^{-n}$ satisfying $L = W\partial W^{-1} (\partial =
\partial_{x}, x\sim t_{1}$ - some authors use $x + t_{1}$ in the $t_{1}$
position).  An operator M (or G) is defined via
\medskip
$$\leqno(4.1)\quad M = W(\sum^\infty_{1}kt_{k}\partial^{k-1})W^{-1} =
G + \sum^\infty_{2}kt_{k}L^{k-1};$$
$$G = WxW^{-1}; [L,M] = [L,G] = 1$$
\medskip
\noindent The operator M is connected to the wave function
$We^{\xi} = (1 + \sum^\infty_{1}w_{n}\lambda^{-n})e^{\xi} = \\
w, \xi = \sum^\infty_{1}t_{n}\lambda^{n},$ via $\partial_{\lambda}w =
Mw$, and M arises in the study of nonisospectral symmetries for
example (see [3;4;7;11;12;32]).
Further
\medskip
$$\leqno(4.2)\quad M = \sum^\infty_{1}jt_{j}L^{j-1} + \sum^\infty_{1}
V_{j+1}L^{-j-1}; V_{j+1} = -js_{j}(-\tilde{\partial})log(\tau);$$
$$G = x + \sum^\infty_{1}V_{j+1}L^{-j-1}$$
\medskip
\noindent where $\tau$ is the tau function (cf. [3;4;6-8;11;42;43]
for discussion).  In particular one should recall that $(\partial_{n} =
\frac{\partial}{\partial t_{n}})$
\medskip
$$\leqno(4.3)\quad X(\lambda)\tau = w\tau; X^{*}(\lambda)\tau =
w^{*}\tau; X(\lambda) = e^{\sum t_{n}\lambda^{n}}
e^{-\sum\frac{\partial_{n}}{n\lambda^{n}}};$$
$$X^{*}(\lambda) = e^{-\sum t_{n}\lambda^{n}}e^{\sum\frac{\partial_{n}}
{n\lambda^{n}}}; w^{*} = W^{*-1}e^{-\xi} = (1 + \sum^\infty_{1}
w^{*}_{i}\lambda^{-i})e^{-\xi}$$
\medskip
\noindent Now to go to dispersionless KP (i.e. quasiclassical limits)
one writes $t_{n}\to \epsilon t_{n} = T_{n}, x = x_{1}\to \epsilon x = X,
\partial_{n}\to \epsilon\frac{\partial}{\partial T_{n}} =
\epsilon\partial_{n},$ with
\medskip
$$\leqno(4.4)\quad L_{\epsilon} = \epsilon\partial + \sum^\infty_{1}
u_{n+1}(\epsilon,T)(\epsilon\partial)^{-n};
w\sim e^{\frac{1}{\epsilon}S(\lambda,T) + O(1)};
\tau\sim e^{\frac{1}{\epsilon^{2}}F(T) + O(\frac{1}{\epsilon})}$$
\medskip
\noindent The latter expression for $\tau$ is really an "ansatz" (cf.
[3;4;51;52]) which is verified in examples (see e.g. [27]).  One
assumes that $u_{n+1}(\epsilon,T)\to \tilde{u}_{n+1}(T)$ as
$\epsilon\to 0$ and we call this again $u_{n+1}(T)$; we omit a
discussion of the philosophy of slow variables, averaging, etc.
(cf. [3;4;13;18;25;42;43]).  We recall that the standard embellished
(with M) KP hierarchy equations $\partial_{n}L = [B_{n},L],
\partial_{n}M = [B_{n},M], B_{n} = L^{n}_{+}, Lw = \lambda w, \partial_{n}w
= B_{n} w,$ become
\medskip
$$\leqno(4.5)\quad [L,M] = \epsilon; \epsilon\partial_{n}w = B_{n}w;
w = \frac{\tau(\epsilon,T_{n} - \frac{\epsilon}{n\lambda^{n}})
e^{\sum^\infty_{1}\lambda^{n}(\frac{T_{n}}{\epsilon})}}{\tau(\epsilon,T)}$$
\medskip
\noindent etc.  Write now $P = \partial S = \frac{\partial S}
{\partial T_{1}}, (T_{1} = X)$ and $\epsilon^{i}\partial^{i} w\to
P^{i}w$ as $\epsilon\to 0$.  Then $Lw = \lambda w$ becomes
\medskip
$$\leqno(4.6)\quad \lambda = P + \sum^\infty_{1}u_{n+1}P^{-n};
P = \lambda - \sum^\infty_{1} P_{i}\lambda^{-i}$$
\medskip
\noindent (the latter equation being simply the inversion of the first).
{}From $B_{n}w = \sum^{n}_{0}b_{n,m}(\epsilon\partial)^{m}w$ one gets
(with some abuse of notation)
\medskip
$$\leqno(4.7)\quad B_{n}\to {\cal B}_{n} = \sum^{n}_{0}b_{n,m}P^{m} =
\lambda^{n}_{+}; \partial_{n}S = {\cal B}_{n}(P); \partial_{n}P =
\hat {\partial}{\cal B}_{n}(P) =$$
$$\partial_{X}{\cal B}_{n} +\partial_{P}{\cal B}_{n}(\frac{\partial P}
{\partial X}); M\to {\cal M} =\sum^\infty_{1}nT_{n}\lambda^{n-1}
 + \sum^\infty_{1}V_{n+1}\lambda^{-n-1}$$
\medskip
\noindent and writing $\{A,B\} = \partial_{P}A\partial_{X}B -
\partial_{X}A\partial_{P}B$ there results
\medskip
$$\leqno(4.8)\quad \partial_{n}\lambda = \{{\cal B}_{n},\lambda\};
\partial_{n}{\cal M} = \{{\cal B}_{n},{\cal M}\}; \{\lambda,{\cal M}\} = 1;
\partial_{n}S = {\cal B}_{n}; \partial_{\lambda}S = {\cal M};$$
$$S = \sum^\infty_{1}T_{n}\lambda^{n} + \sum^\infty_{1}S_{i+1}\lambda^{-i};
\partial S_{n+1} = -P_{n}; V_{n+1} = -nS_{n+1} = \partial_{n}log(\tau)
= \partial_{n}F$$
\medskip
\noindent Then one defines $\tau^{dKP}  = e^{F}, F = F(X,\hat {T}).$\\
\indent In [3;4] we showed how this framework is related to the Hamilton-
Jacobi theory of [32] for dKP.  In particular it is important to rescale
the $T_{n}$ to $T'_{n} = nT_{n},\partial_{n}\to n\frac{\partial}
{\partial {T'}_{n}}$ (this rescaling is also involved in the Landau-Ginsburg
equation and connections to gravity (cf. [2-4;13;25;42;43;49]).
Then one gets
$$\leqno(4.9)\quad \partial'_{n}S = \frac{\lambda^{n}_{+}}{n};
\partial'_{n}\lambda = \{{\cal Q},\lambda\}; {\cal Q}_{n} = \frac{{\cal B}_{n}}
{n}; \partial'_{n}P = \hat {\partial}{\cal Q}_{n} = \partial{\cal Q}_{n}
+ \partial_{P}{\cal Q}_{n}\partial P$$
\medskip
\noindent Further with $(P,X,T'_{n}), n\geq {2}$, as Hamiltonian variables,
$P = P(X,T'_{n}), -{\cal Q}_{n}\\ = -{\cal Q}_{n}(P,X,T'_{n}) =$
Hamiltonian, there results
\medskip
$$\leqno(4.10)\quad \dot {P}_{n} = \frac{dP}{dT'_{n}} =
\partial{\cal Q}_{n}; \dot {X}_{n} = - \partial_{P}{\cal Q}_{n}$$
\medskip
\noindent Setting $PdX + {\cal Q}_{n}dT'_{n} = -\hat {\xi}d\lambda -
K_{n}dT'_{n} + d\tilde {S}, K_{n} = -R_{n} = -\frac{\lambda^{n}}{n},$
with $\tilde {S}_{\lambda} = \hat {\xi}, \tilde {S}_{X} = P,
\partial'_{n}\tilde {S} = {\cal Q}_{n} - R_{n},$ we have action-angle
variables $(\lambda,-\hat {\xi})$ where
\medskip
$$\leqno(4.11)\quad \hat {\xi} = S_{\lambda} - \sum^\infty_{2}T'_{n}
\lambda^{n-1} = lim WxW^{-1} = lim G = {\cal M} - \sum^\infty_{2}
T'_{n}\lambda^{n-1}$$
\medskip
\noindent
Let us also remark that, posing $w^{*} = exp(\frac{S^{*}}{\epsilon} +
O(1)),$ one obtains as above $(w^{*} = e^{-\xi}(\frac{\tau_{+}}{\tau}))$
\medskip
$$\leqno(4.12)\quad log(w^{*})\sim\frac{S^{*}}{\epsilon} + O(1)\sim
-\frac{1}{\epsilon}\sum^\infty_{1}T_{n}\lambda^{n} +
\frac{1}{\epsilon}\sum^\infty_{1}\frac{\partial_{n}F}{n\lambda^{n}}$$
\medskip
\noindent Consequently $S^{*} = -S.$\\
\vspace{3mm}
\indent We recall next that tau functions are generated via B$\ddot a$cklund
type actions from the Clifford group (cf. [8;9;19]).  In the bosonic
picture this translates into generation by vertex operators (cf. (4.3))
\medskip
$$\leqno(4.13)\quad X(\lambda,\zeta) = \frac{(\lambda - \zeta)}{\zeta}X^{*}
(\zeta)X(\lambda) = \frac{(\lambda - \zeta)}{\lambda}X(\lambda)X^{*}(\zeta)$$
$$= e^{\xi(x,\lambda) - \xi(x,\zeta)}e^{-\xi(\tilde {\partial},\lambda^{-1})
+ \xi(\tilde{\partial},\zeta^{-1})}$$
\medskip
\noindent $(\tilde{\partial} = (\partial_{1},\frac{1}{2}\partial_{2},...))$.
Write now e.g. $\xi'(x,\lambda) = \sum^\infty_{2}x_{n}\lambda^{n}$
with $\lambda\in S^{1}$ so via (2.13)
\medskip
$$\leqno(4.14)\quad X(\lambda) = e^{x\lambda + \xi'(x,\lambda)}
e^{-\bar{\lambda}\partial - \xi'(\tilde{\partial},\bar{\lambda})} =
e^{x\lambda}e^{\xi'(x,\lambda)}e^{-\bar{\lambda}\partial}
e^{-\xi'(\tilde{\partial},\bar{\lambda})}$$
$$= e^{\xi'(x,\lambda)}e^{-\xi'(\tilde{\partial},\bar{\lambda})}
e^{-\frac{1}{2}[x,\partial]}e^{x\lambda - \bar{\lambda}\partial} =
e^{\frac{1}{2}}e^{\xi'(x,\lambda)}e^{-\xi'(\tilde{\partial},\bar{\lambda})}
\hat{D}(\lambda)$$
\medskip
\noindent where $\hat{D}(\lambda) = e^{\lambda a^{\dagger} - \bar{\lambda}a}$
(cf. (A.6) where $D(z) = exp(za^{\dagger}_{h} - za_{h})$ and recall
$\frac{\lambda}{\epsilon\sqrt{2}}\sim z$ from (3.6)).  Similarly for
$\zeta\in S^{1}$
\medskip
$$\leqno(4.15)\quad X^{*}(\zeta) = e^{-\xi'(x,\zeta)}e^{\xi'(\tilde{\partial},
\bar{\zeta})}e^{\frac{1}{2}}\hat{D}(-\zeta)$$
\medskip
\noindent This leads to
\medskip
$$\leqno(4.16)\quad  X(\lambda,\zeta) = \frac{(\lambda - \zeta)}{\lambda}
\tilde{X}(\lambda,\zeta)\hat{D}(\lambda)\hat{D}(-\zeta);$$
$$\tilde{X}(\lambda,\zeta) = e^{\xi'(x,\lambda)}e^{-\xi'(\tilde{\partial},
\bar{\lambda})}e^{-\xi'(x,\zeta)}e^{\xi'(\tilde{\partial},\bar{\zeta})}e$$
\medskip
\noindent One could continue and pass any finite number of Bose operators
$a_n \sim \partial_n,\,a_n^{\dagger}\sim x_n$ to the right (but we do
not know at this time how to construct the limiting geometrical object
$\sim$ dispersionless Grassmannian).  Here we
simply want to distinguish the $(x,\partial)$ variables
as determining a phase space in the spirit of dispersionless limits
(cf. (4.4)-(4.11)) and hence we will treat $(x,\partial)$ as special and
think of the other $x_{n}\sim t_{n}$ as time parameters
\\[3mm]\indent One can also think of multisoliton tau functions of
the form $(X^{2}(\lambda,\zeta) = 0)$
\medskip
$$\leqno(4.17)\quad \tau_{N} = \prod^{N}_{1}(1 + a_{j}X(\lambda_{j},
\zeta_{j})) \cdot 1$$
\medskip
\noindent in the bosonic picture, and more generally one considers a limiting
procedure with $N\to {\infty}$ (cf. [9;19]).  It is folkloric
that such multisoliton constructions will be dense in some sense
but we will not try to clarify that here.  Such $\tau_{N}$ arise from
a construction $\tau(x,g) = <0|exp(H(x))g|0>$ for $H(x) = \sum^\infty_{1}
x_{k}J_{k}, J_{k}\to {\partial_{k}}, g = exp(\sum^{N}_{1}
\frac{(\lambda_{j} - \zeta_{j})}{\zeta_{j}}\psi(\lambda_{j})
\psi^{*}(\lambda_{j}))$ where the $\psi,\psi^{*}$ operators are
built up from free fermion operators and need not concern us here
(cf. [8;9;19]).  For the quasiclassical or dispersionless situation
one must insert $\epsilon$ at appropriate places to arrive e.g. at
(cf. [42;43])
\medskip
$$\leqno(4.18)\quad \tau^{\epsilon}_{N}\sim \prod^{N}_{1}
(1 + \frac{a_{j}}{\epsilon}X(\frac{T}{\epsilon},\lambda_{j},\zeta_{j}))
\cdot 1$$
\medskip
\noindent and we will assume $\lambda_{j},\zeta_{j}\in S^{1}$ in
what follows.  The vacuum vector $1\in {\cal B}$ = polynomial Fock space
corresponds to the boson representation and we can think of
$\tau^{\epsilon}_{N}\in \hat {{\cal B}}$ = Fock space based on
$(\partial,x)$ with the $x_{n}\sim t_{n}$ or $T_{n}$ variables as
parameteres.  Then the vacuum $1 = |0>$ can also be represented in terms
of peaked states by our linking procedure and we can look at a phase
space $\sim$ coadjoint orbit based on coordinates p,q as in section 2.
The question then is to relate this to the phase space based on X,P or
$\lambda,\hat {\xi}$ obtained in (4.4)-(4.11).
\\[.5cm]\noindent {\bf REMARK 4.1} $\,$ Let us note that for $\lambda\in
S^{1}, \lambda a^{\dagger} - \bar {\lambda} a$ in (3.5) (cf. Theorem 3.1)
has the form $(\frac{i\sqrt{2}}{\epsilon})(p\hat{Q} - p\hat{P})$ with
\medskip
$$\leqno(4.19)\quad \frac{1}{\epsilon\sqrt{2}}(\lambda a^{\dagger} -
\bar {\lambda} a) = \zeta a^{\dagger}(\epsilon) - \bar {\zeta} a(\epsilon)
= \frac{i}{\epsilon^{2}}(p\hat{Q} -q\hat{P})$$
\medskip
\noindent so $\lambda a^{\dagger} - \bar {\lambda} a$ lies inbetween a
classical operator $i(\tilde {p}\hat{Q} - \tilde {q}\hat{P})$ and a
coherent state generator $\frac{i}{\epsilon^{2}}(\tilde{p}\hat{Q} -
\tilde{q}\hat{P}),$ where $\tilde{p} = \sqrt{2}Im(\lambda)$ and
$\tilde{q} = \sqrt{2}Re(\lambda) ((\tilde{p},\tilde{q}) = \sqrt{2}(p,q))$
Now via section 2 one has for $z = h^{-\frac{1}{2}}\alpha\sim
\frac{1}{\epsilon\sqrt{2}}(\xi + i\pi)$
\medskip
$$\leqno(4.20)\quad <z|(\hat{Q} - \xi)(\hat{P} - \pi)|z>\to 0$$
\medskip
\noindent so $\hat{Q}\to \xi$ and $\hat{P}\to \pi$ (note $(\xi,\pi)\not =$
(q,p)).  Observe that if we write in (3.4) $a\sim \epsilon\partial_{X}$
and $a^{\dagger}\sim \frac{X}{\epsilon}$ then
\medskip
$$\leqno(4.21)\quad \frac{X}{\epsilon} = \frac{(\hat{Q} -i\hat{P})}
{\epsilon\sqrt{2}}; \frac{\epsilon^{2}\partial_{X}}{\epsilon} =
\frac{1}{\epsilon\sqrt{2}}(\hat{Q} + i\hat{P})$$
\medskip
\noindent Formally it follows that $X\to \frac{(\xi - i\pi)}{\sqrt{2}}$
and $\epsilon^{2}\partial_{X}\to \frac{(\xi + i\pi)}{\sqrt{2}}$ (cf.
(4.33) for further validation).  Thus phase space variables $\xi,\pi$
coming from z can be compared to the scaled X variable, which arises
in the dispersionless KP situation.  We emphasize here that $(\pi,\xi)\not =
(p,q),$ where we stipulate that $(p,q)\sim \zeta$ now.
\\[5mm]\noindent {\bf REMARK 4.2} $\,$ Assume we have built up a tau
function as in (4.18) and think of it in $\hat {{\cal B}}.$  One can
carry all $\partial_{n} (n\geq 2)$ to the right to work on the 1 vacuum
of ${\cal B}$ for example so that there remains a sum of terms = functions
of $(T_{n},\epsilon), n\geq 2,$ times operators $\hat{D}_{\epsilon}
(\lambda_{i})\hat{D}_{\epsilon}(-\zeta_{i})\frac{\lambda_{i} -
\zeta_{i}}{\lambda_{i}}$ as in (4.16) (see below for $\hat{D}_{\epsilon}).$
Note how $\epsilon$ arises in (4.18) so analogously to (4.16) we would have
terms
\medskip
$$\leqno(4.22)\quad \frac{a_{j}}{\epsilon}\tilde{X}(\frac{\hat{T}}{\epsilon},
\lambda_{j},\zeta_{j})\frac{(\lambda_{j} - \zeta_{j})}{\lambda_{j}}
\hat{D}_{\epsilon}(\lambda_{j})\hat{D}_{\epsilon}(-\zeta_{j});$$
$$\hat{D}_{\epsilon}(\lambda_{j}) = e^{\frac{1}{\epsilon}\lambda_{j}X -
\epsilon\partial_{X}\bar{\lambda}_{j}} = e^{\frac{i}{\epsilon}
(\tilde{p}_{j}\hat{Q} - \tilde{q}_{j}\hat{P})}$$
\medskip
\noindent One can use $a(\epsilon),\,a^{\dagger}(\epsilon)$, based on
$(\frac{X}
{\epsilon},\epsilon\partial_X)\equiv (x,\partial)$ interchangeably
with $(x,\partial)$, consistent with (4.21).  The
$\hat{D}_{\epsilon}(\lambda)$ in (4.22) lie
inbetween a classical operator and a coherent state generator and we will
refer to them  as semiclassical or
quasiclassical operators.
\\[5mm]\noindent {\bf THEOREM 4.3.} $\,$ Let $\hat{D}_{\epsilon}(\lambda)$
be semiclassical.  Then one can write
\medskip
$$\leqno(4.23)\quad <z|\hat{D}_{\epsilon}(\lambda)|z>\sim e^{-\frac{1}{2}}
e^{[(\frac{i}{\epsilon})(\tilde{p}\xi - \tilde{q}\pi)]}$$
\medskip
\noindent where $z\sim h^{-\frac{1}{2}}\alpha = h^{-\frac{1}{2}}
\frac{(\xi + i\pi)}{\sqrt{2}}, h = \epsilon^{2}, \tilde{p} =
\sqrt{2}p, \tilde{q} = \sqrt{2}p, p = Im(\lambda), q = Re(\lambda),
\lambda a^{\dagger}(\epsilon) - \bar{\lambda}a(\epsilon)) =
\frac{i}{\epsilon}(\tilde{p}\hat{Q} - \tilde{q}\hat{P}), \hat{Q} =
\hat{q}_{h} = \epsilon \hat{q}; \hat{P} = \hat{p}_{h} = \epsilon\hat{p},$
and $\hat{D}_{\epsilon}(\lambda) = exp(\lambda a^{\dagger}(\epsilon) -
\bar{\lambda}a(\epsilon)).$
\\[3mm]\indent{\it Proof:} $\,$ First consider (2.11) in the form
\medskip
$$\leqno(4.24)\quad <z|\hat{q} - \frac{\xi}{\sqrt{h}}|z> = <z|
\frac{(\hat{Q} - \xi)}{\sqrt{h}}|z> = <0|\hat{q}|0>$$
$$\Longrightarrow <z|\hat{Q} - \xi)|z> = \sqrt{h}
<0|\hat{q}|0>\to 0$$
\medskip
\noindent Similarly $<z|(\hat{P} - \pi)|0> = {h}^{\frac{1}{2}}
<0|\hat{p}|0>\to 0.$  Here $z = h^{-\frac{1}{2}}\alpha =
h^{-\frac{1}{2}}\frac{(\xi + i\pi)}{\sqrt{2}}.$  Consider $<z|
(\lambda a^{\dagger} - \bar{\lambda}a)|z>$ for general z.  Write
$\frac{\lambda}{\epsilon\sqrt{2}} = \zeta$ so
\medskip
$$\leqno(4.25)\quad \zeta a^{\dagger}(\epsilon) - \bar{\zeta}
a(\epsilon) = \frac{i}{\epsilon^{2}}(p\hat{Q} - q\hat{P});
\lambda a^{\dagger}(\epsilon) - \bar{\lambda}a(\epsilon) =
\frac{i}{\epsilon}\sqrt{2}(p\hat{Q} - q\hat{P})$$
\medskip
\noindent Let $z\sim h^{-\frac{1}{2}}\alpha$ as above, based on
$(\xi,\pi) ((\xi,\pi)\not = (q,p))$.  We know
\medskip
$$\leqno(4.26)\quad <z|(\hat{Q} - \xi)|z> = \epsilon<0|\hat{q}|0>
\Rightarrow \frac{i}{\epsilon}p\sqrt{2}<z|(\hat{Q} - \xi)|z>$$
$$= ip\sqrt{2}<0|\hat{q}|0>; -\frac{i}{\epsilon}q\sqrt{2}<z|(\hat{P} -
\pi)|z> = -iq\sqrt{2}<0|\hat{p}|0>$$
\medskip
\noindent Consequently
\medskip
$$\leqno(4.27)\quad <z|(\lambda a^{\dagger}(\epsilon) - \bar{\lambda}
a(\epsilon) -(\frac{i\sqrt{2}}{\epsilon})(p\xi -q\pi)|z> =$$
$$\frac{i\sqrt{2}}{\epsilon}<z|(p\hat{Q}-q\hat{P}) - (p\xi-q\pi)|z> =
i\sqrt{2}(p<0|\hat{q}|0> - q<0|\hat{p}|0>)$$
\medskip
\noindent We can assume $<0|\hat{q}|0> = <0|\hat{p}|0> = 0$ (see below) so
\medskip
$$\leqno(4.28)\quad <z|(\lambda a^{\dagger}(\epsilon) - \bar{\lambda}
a(\epsilon))|z> = \frac{i\sqrt{2}}{\epsilon}(p\xi - q\pi)$$
\medskip
\noindent where p,q come from $\lambda$ and $\pi,\xi$ from z.  Note
for $z = 0, \xi = 0$ we have $<0|\hat{Q} - 0|0>\to 0 \Rightarrow
<0|\hat{q}|0> = 0$ etc.  This suggests that functions like $\hat{D}_
{\epsilon}(\lambda) = exp(\lambda a^{\dagger}(\epsilon) - \bar{\lambda}
a(\epsilon))$ should also have limit expressions.  In particular
\medskip
$$\leqno(4.29)\quad \langle z|\hat{D}_{\epsilon}(\lambda)|
z \rangle \sim ce^{(\frac{i\sqrt{2}}{\epsilon})(p\xi - q\pi)}$$
\medskip
\noindent should be valid for a suitable c (see below).  For confirmation
we recall first from [47] that for classical operators $\hat{A} (A_{h}
= <z|\hat{A}|z>_{h})$
\medskip
$$\leqno(4.30)\quad (AB)_{h} = \int d\mu(|\langle u|u'\rangle|^{2})
[\frac{<u|\hat{A}|u'>}{<u|u'>}\frac{<u'|\hat{B}|u>}{<u'|u>}]$$
$$ + o(1) = A_{h}(u)B_{h}(u) + o(1)$$
\medskip
\noindent  One could
surely base a proof of (4.31) and a determination of c based upon (4.30)
but there is a simpler approach using a theorem from [17].  First note
in (4.21) with $X\to \frac{(\xi - i\pi)}{\sqrt{2}}, \epsilon^{2}\partial_{X}
\to \frac{(\xi + i\pi)}{\sqrt{2}},$ based on section 2 (as in (4.28)),
that ($\to$ in the sense of (4.28))
\medskip
$$\leqno(4.31)\quad \frac{1}{\epsilon}\lambda X - \epsilon\bar{\lambda}
\partial\to \frac{1}{\epsilon\sqrt{2}}[(q+ip)(\xi-i\pi) -(q-ip)
(\xi+i\pi)]$$
$$ = \frac{i\sqrt{2}}{\epsilon}(p\xi -q\pi)$$
\medskip
\noindent and one can use this this as a validation of the identification
in (4.21).  Moreover the arrow $\to$ is unnecessary since in fact
$<z|\hat{Q}|z> = \xi<z|z> = \xi$ etc.  Now to confirm (4.29) we refer
to [17] and recall from section 2, $U(h^{-\frac{1}{2}}\alpha)\sim D(z),
z = h^{-\frac{1}{2}}\alpha$.  For suitable problems (2.1)-(2.2) one
specializes Theorem 2.1 in [17] (cf. also [45]) to $t = 0$ to obtain
(limit in strong operator topology)
\medskip
$$\leqno(4.32)\quad U(h^{-\frac{1}{2}}\alpha)^{\dagger}e^{i[r(\hat{q} -
h^{-\frac{1}{2}}\xi) + s(\hat{p} - h^{-\frac{1}{2}}\pi)]}U(h^{-\frac{1}{2}}
\alpha)\to e^{i(r\hat{q} + s\hat{p})}$$
\medskip
\noindent Since we are at liberty to choose any problem (2.1)-(2.2) giving
peaked states (based on the identification (3.4)) there is no problem
in using Theorem 2.1 of [17].  Then since $\hat{D}_{\epsilon}(\lambda) =
exp(\frac{i}{\epsilon}(\tilde{p}\hat{Q} - \tilde{q}\hat{P})), h\sim
\epsilon^{2}, \hat{Q}\sim \hat{q}_{h},$ etc. we take $r = \tilde{p}$ and $s =
-\tilde{q}$ in (4.32) to obtain
\medskip
$$\leqno(4.33)\quad <z|e^{\frac{i}{\epsilon}[\tilde{p}(\hat{Q} -
\xi) - \tilde{q}(\hat{P} - \pi)]}|z>\to <0|e^{i(\tilde{p}\hat{q} -
\tilde{q}\hat{p})}|0> = c(\tilde{p},\tilde{q})$$
\medskip
\noindent But $i(\tilde{p}\hat{q} - \tilde{q}\hat{p})\sim
\lambda a^{\dagger} - \bar{\lambda}a$ with $a|0> = 0$ so by
(2.13) $c(\tilde{p},\tilde{q}) = e^{-\frac{1}{2}}$ and
\medskip
$$\leqno(4.34)\quad <z|e^{\frac{i}{\epsilon}(\tilde{p}\hat{Q} -
\tilde{q}\hat{P})}|z>\sim e^{-\frac{1}{2}}e^{\frac{i}{\epsilon}(\tilde{p}\xi -
\tilde{q}\pi)}$$
\medskip
\noindent which corresponds to (4.29).  QED
\\[3mm]\indent From the construction (2.16) for classical operators
one sees an immediate generalization for quasiclassical operators.
Thus eliminate the u term in (2.16) and write (cf. (2.14))
\medskip
$$\leqno(4.35)\quad \hat{A} = \int \int dpdq\tilde{f}(p,q)U(hp,hq)$$
\medskip\noindent (e.g. $u = -\frac{1}{2}pq$ removes u and factors of h
in $\tilde{f}$ automatically vanish - or simply integrate out the u
term).  Now $U(hp,hq)\sim exp[i(p\hat{Q} - q\hat{P})]$ so we suggest
that general quasiclassical operators $\hat{A}_{QC}$ can be obtained via
\medskip
$$\leqno(4.36)\quad \hat{A}_{QC} = \int \int d\tilde{p}d\tilde{q}
\tilde{f}(\tilde{p},\tilde{q})\hat{D}_{\epsilon}(\lambda);
\hat{D}_{\epsilon}(\lambda) = e^{\frac{i}{\epsilon}[\tilde{p}\hat{Q}
- \tilde{q}\hat{P}]}$$
\medskip
\noindent In view of (4.23), for this to make sense one would specify that
\medskip
$$\leqno(4.37)\quad <z|\hat{A}_{QC}|z>\sim e^{-\frac{1}{2}} \int \int
e^{\frac{i}{\epsilon}(\tilde{p}\xi - \tilde{q}\pi)}\tilde{f}(\tilde{p},
\tilde{q})d\tilde{p}d\tilde{q}$$
\medskip
\noindent be valid.  Thus the integral should be well defined and the
error term suitably small.  Heuristically we state here (cf. Appendix
B for more structure)
\\[3mm]\indent {\bf COROLLARY 4.4.} $\,$ One can heuristically define a
class of quasiclassical operators via (4.36).
\\[3mm]\indent {\bf REMARK 4.5.} $\,$ We note that the idea of quasiclassical
objects in [42;43] uses a KP based h instead of $\epsilon$, which overlooks
the type of explicit connection to quantum mechanical ideas indicated in the
present paper. Physically one can perhaps think of the $\epsilon\sim
\sqrt{h}$ smoothing of quantum fluctuations as related to an interaction
between dispersionless limits and weak solutions in fluid dynamics.
\\[3mm]\indent {\bf REMARK 4.6.} $\,$ Theorem 4.3 shows that phase space
calculations based on $\hat{D}_{\epsilon}(\lambda)$ have an asymptotic
character as in (4.23) and this agrees (up to a constant $c(\tilde{p},
\tilde{q})$ with a direct calculation based on $\frac{1}{\epsilon}X$ and
$\epsilon^{2}\partial$ as in (4.21) and (4.31) (cf. (4.22)).  The soliton
calculation, not using $\hat {{\cal B}}$, would pass $\partial_{X}$ to the
right where it would act on 1, thus eliminating it's contribution, and this
would change the exponential factor corresponding to (4.23).  However
(4.23) can be recast via (4.31) in terms of $\frac{1}{\epsilon}X$ and
$\epsilon^{2}\partial\sim i{\cal P},$ (cf. Remark 4.7), and eliminating the
$i{\cal P}$ contribution one obtains the equivalent soliton calculation.
Note that the soliton calculation approach does
not a priori inject S via ${\cal P}\sim P = \partial S$ into the equations;
S, and thence P, appears as a result of the calculations.
In this connection we note also that a finite product of terms $\hat{D}_
{\epsilon}(\lambda_{j})\hat{D}_{\epsilon}(-\zeta_{j})$ arising out of (4.18)
for example leads to (cf. (4.22), (2.13))
\medskip
$$\leqno(4.38)\quad \prod(\frac{a_{j}}{\epsilon})\frac{(\lambda_{j} -
\zeta_{j})}{\lambda_{j}}\tilde{X}(\frac{\hat{T}}{\epsilon},\lambda_{j},
\zeta_{j})\prod \hat{D}_{\epsilon}(\lambda_{j})\hat{D}_{\epsilon}
(-\zeta_{j})$$
$$= \tilde{\phi}(\lambda_{j},\zeta_{j},\hat{T},\epsilon)\hat{D}_{\epsilon}
(\sum \lambda_{j} - \sum  \zeta_{j})$$
\medskip
\noindent where $\tilde{\phi} = \phi exp(-iIm\sum \lambda_{j}\bar{\zeta}_{j})
\cdot\Xi,\,\, \Xi = exp[(\lambda_{1}-\zeta_{1})\sum^{N}_{2}({(\bar{\lambda_{j}}
-\bar{\zeta_{j}})}
+ (\lambda_{2}-\zeta_{2})\sum^{N}_{3}{(\bar{\lambda_{j}}-\bar{\zeta_{j}})}
+ ... + (\lambda_{N-1}-\zeta_{N-1}){(\bar{\lambda_{N}}-\bar{\zeta_{N}})]}$.
Thus all terms (4.38) are expressed as operators in $\hat{{\cal B}}$
via $\hat{D}_{\epsilon}(\Lambda-Z), \Lambda = \sum\lambda_{j}, Z =
\sum\zeta_{j},$ and the estimates obtained via (4.23) apply.  It follows
that the asymptotic estimates $\tau\sim exp(\frac{1}{\epsilon^{2}}F)$ for
KP based on quasiclassical soliton calculations are unchanged and our approach
gives a geometrical background for the quasiclassical soliton procedure.
Conceptually one can avoid the original scaling step in x by arguing via
$a\sim \partial$ and $a^{\dagger}\sim x$.  Thus one has given various
$\hat {D}(\lambda) = exp(\lambda a^{\dagger} - \bar {\lambda}a)$ and
the insertion of $\epsilon$ can be thought of as a way of introducing
peaked states and thence coadjoint orbit variables.
\\[3mm]\indent {\bf REMARK 4.7.} $\,$ Let us write $\epsilon^{2}\partial_{X}
\sim i{\cal P}$ and rephrase (4.31) as
\medskip
$$\leqno(4.39)\quad X\sim\frac{(\xi - i\pi)}{\sqrt{2}}; i{\cal P}\sim
\frac{(\xi + i\pi)}{\sqrt{2}}; \xi\sim \frac{(X + i{\cal P})}{\sqrt{2}};
i\pi\sim\frac{(i{\cal P} - X)}{\sqrt{2}}$$
\medskip
\noindent Here $(\xi,\pi)\sim z, (q,p)\sim\lambda,$ and $(X,{\cal P})\sim
a(\epsilon)$.  From section 2 the z coordinates appear in Poisson brackets
$\{f,g\} = \zeta_{3}(f_{\pi}g_{\xi} - f_{\xi}g_{\pi})$ while from [3;4]
Poisson brackets for X and $P = \partial S$ are eventually defined via
$\{A,B\} = \partial_{P}A\partial_{X}B - \partial_{X}A\partial_{P}B$ so that
$\{P,X\} = 1$.  From (4.39)
\medskip
$$\leqno(4.40)\quad \partial_{\pi} = \frac{1}{\sqrt{2}}( \partial_{{\cal P}} -
i\partial_{X}); \partial_{\xi} = \frac{1}{\sqrt{2}}(\partial_{X} -
i\partial_{{\cal P}}); f_{\pi}g_{\xi} - f_{\xi}g_{\pi} =
f_{{\cal P}}g_{X} - f_{X}g_{{\cal P}}$$
\medskip
\noindent so one has a natural identification of P and ${\cal P}$ for
$\zeta_{3} = 1$.  Note also that (4.10) becomes then $\dot {\xi} =
-\partial_{\pi}{\cal Q}_{n}$ and $\dot {\pi} = \partial_{\xi}{\cal Q}_{n}.$
\\[3mm]\indent {\bf REMARK 4.8.} $\,$ There is actually no restriction
of the form $(\lambda_{j}, \zeta_{j})\in S^{1}$ as long as we keep our
coherent states $|z>$ based as before on unitary operators.  Thus e.g.
in (4.38) replace $\hat {D}_{\epsilon}(\lambda_{j})$ and $\hat
{D}_{\epsilon}(-\zeta_{j})$ arising out of (4.18) by $exp(\frac{1}{\epsilon}
\lambda_{j}x - \epsilon \partial \lambda_{j}^{-1}) = exp(\lambda_{j}
a^{\dagger}(\epsilon) - \lambda_{j}^{-1}a(\epsilon)) = \tilde {D}
(\lambda_{j},-\lambda_{j}^{-1})$.  Then
$\tilde {D}(\lambda_{j},-\lambda_{j}^{-1})\tilde {D}(-\zeta_{j},
\zeta_{j}^{-1}) = exp(\frac{1}{2}(\zeta_{j}\lambda_{j}^{-1} -
\lambda_{j}\zeta_{j}^{-1}))\tilde {D}(\lambda_{j}-\zeta_{j},
\zeta_{j}^{-1} - \lambda_{j}^{-1})$ so finite products as in (4.38) become
\medskip
$$\leqno(4.41)\quad \phi(\lambda_{j},\zeta_{j},\hat {T},\epsilon)
\prod e^{\frac{1}{2}(\zeta_{j}\lambda_{j}^{-1}-\lambda_{j}\zeta_{j}^{-1})}
\tilde {D}(\lambda_{j}-\zeta_{j},\zeta_{j}^{-1} -
\lambda_{j}^{-1}) =$$
$$\hat {\phi}(\lambda_{j},\zeta_{j},\hat{T},\epsilon)\hat{D}(\sum
(\lambda_{j}-\zeta_{j}),\sum (\zeta_{j}^{-1} - \lambda_{j}^{-1})) =
\hat {\phi}\tilde {D}(\mu,\nu)$$
\medskip
\noindent For specific choices of $(\lambda_j,\zeta_j)$ one could
use (4.23), etc.
\\[3mm]\indent {\bf REMARK 4.9.} $\,$  One can
deal with coherent state manifolds based on $gl(\infty)$
in a general context suggested in [34;35].
Given e.g. imaginary time coordinates $t_{n}$, Lagrange equations can be
developed via path integrals, etc. (cf. [26;34;35;37;41]) but a
geometrical transition to quasiclassical or dispersionless limits via
collapse of coherent states to coadjoint orbits seems unclear at this time.
We expect that some variation on the geometric ideas indicated here via
the connection to quantum mechanics should also apply in the general
soliton situation.  That is, the coherent state manifold should collapse
onto coadjoint orbits, and the corresponding K$\ddot a$hler structures
should be related, etc.  We have not yet made this all explicit
however.
\\[3mm]\indent {\bf REMARK 4.10.} $\,$ Let us be explicit about what
has been accomplished here.  We start basically with $\hat {D}(\lambda)\cdot 1
=
exp(\lambda x - \bar {\lambda}\partial) \cdot 1 = exp(\lambda X)$ which
becomes $\hat {D}_{\epsilon}(\lambda)\cdot 1$ or $exp(\frac{1}{\epsilon}
\lambda x)$ by scaling.  On the other hand
given $e^{\lambda a^{\dagger} - \bar {\lambda}a}$ ($a^{\dagger} \sim x,
a\sim \partial$), we insert $\epsilon$ to generate peaked states via
$a = a(\epsilon) = \frac{1}{\epsilon\sqrt{2}}(\hat {Q} + \epsilon^{2}
\partial_{Q}), a^{\dagger} = a^{\dagger}(\epsilon) = \frac{1}{\epsilon\sqrt{2}}
(\hat {Q} - \epsilon^{2}\partial_{Q})$ as in (3.4)-(3.5).  Then Theorem
4.3 applies and one obtains $<z|\hat {D}_{\epsilon}(\lambda)|z>\sim
e^{-\frac{1}{2}}exp[\frac{i}{\epsilon}(\tilde {p}\xi - \tilde {q}\pi)]
= \hat {d}(\epsilon,\lambda,\xi,\pi)$ as the quasiclassical object
associated with the operator $\hat {D}_{\epsilon}(\lambda).$  Then via
(4.21), (4.31), (4.39) one creates an X variable (incidentally the same
as X obtained by scaling) and rewrites $\hat {d}(\epsilon,\lambda,
\xi,\pi)$ in terms of X and ${\cal P}$.  In the situation where $\hat {D}
(\lambda)$ acts on 1 we need only concern ourselves with $\check {d}
(\lambda) = e^{\lambda a^{\dagger}}$ in which case the corresponding
quasiclassical object $\check {d}(\epsilon,\lambda,\xi,\pi)$ becomes
$\check {d}(\epsilon,\lambda,X) = e^{\frac{1}{\epsilon}\lambda X}$
(cf. also (4.30)).  This may seem like a lot of work to go from
$e^{\frac{1}{\epsilon}\lambda X}$ to $e^{\frac{1}{\epsilon}\lambda X}$
but conceptually we have eliminated the ideas of averaging or scaling
the x variable (via fast and slow variables, etc.); such notions have
been replaced by a more geometrical construction. Evidently one may
apply such techniques to any finite number of $t_{n}, \partial_{n}$ as
well, and this could provide a background structure for some kind of
limiting geometrical object (Grassmannian) alluded to in Remark 4.9.
In the event that the $(\xi,\pi)$ phase space is based on a
constrained region (e.g. $S^{1}$) one has recourse to peaked state
constructions on such regions (cf. [10]).
\\[1cm]\noindent {\bf 5. PHASE SPACES, ACTION, AND WKB.}
\\[.5cm]\indent Let us put some material from [3;4] (cf. (4.1)-(4.12))
in a broader perspective and indicate at the same time some calculations
in the dispersionless theory.  We recall from [3;4] and section 4
\medskip
$$\leqno(5.1)\quad \tilde{S} = S - \sum^\infty_{2}T_{n}\lambda^{n} =
X\lambda - \sum^\infty_{1}(\frac{\partial_{n}F}{n})\lambda^{-n} =$$
$$-\int_{X}^{\infty} [P(X',\lambda,\hat{T}_{n}) - \lambda]dX'
+ \lambda X; \frac{\partial_{n}F}{n} = -\int_{X}^{\infty} P_{n}dX';
P - \lambda = -\sum^\infty_{1} P_{j}\lambda^{-j}$$
\medskip
$$\leqno(5.2)\quad \hat{\xi} = \tilde{S}_{\lambda}; \tilde{S}_{X} =
P = S_{X}; \partial'_{n}\tilde{S} = {\cal Q}_{n} - R_{n}; {\cal Q}_{n} =
\frac{1}{n}{\cal B}_{n} = \partial'_{n}S;$$
$$T'_{n} = nT_{n}; R_{n} = \frac{1}{n}\lambda^{n}; S_{\lambda} =
{\cal M}; K_{n} = -R_{n}$$
\medskip
\noindent Now for $H_{n} = -{\cal Q}_{n}$ we can write
\medskip
$$\leqno(5.3)\quad PdX - H_{n}dT'_{n} = -\hat{\xi}d\lambda
-K_{n}dT'_{n} + d\tilde{S}$$
\medskip
\noindent which reveals $\tilde{S}$ as a generating function of type
$\tilde{S}(X,\lambda,\hat{T}_{n}) = F_{1}(X,\lambda,\hat{T}_{n})$
for a canonical transformation $(X,P)\to (\lambda,
-\hat{\xi})$, and $(\lambda,-\hat{\xi})$ are action-angle variables with
\medskip
$$\leqno(5.4)\quad {d\lambda \over dT'_{n}} = \dot{\lambda}_{n} = 0;
{d\hat{\xi} \over dT'_{n}} = \dot{\hat{\xi}}_{n} = \partial_{\lambda}
K_{n}$$
\medskip
\noindent (and $\dot{P}_{n} = \partial{\cal Q}_{n}$ with $\dot{X}_{n} =
-\partial_{P}{\cal Q}_{n})$.  Recall that there are two main types of
generating functions $F_{1}(q,Q,t)$ and $F_{2}(q,P,t)$ for canonical
transformations $(q,p)\to (Q,P)$ satisfying $p\dot{q} - H = P\dot{Q}
- K + dF$ where H and K are Hamiltonians.  One has in particular $K = H +
\partial_{t}F, \dot{P} = -H_{q}, \dot{q} = H_{P}, \dot{P} = -K_{Q},$
and $\dot{Q} = K_{P}$ in both cases and
\medskip
$$\leqno(5.5)\quad p = \partial_{q}F_{1}; P = -\partial_{Q}F_{1};
p = \partial_{q}F_{2}; Q = \partial_{P}F_{2}$$
\medskip
\noindent (thus $-\hat{\xi} = -\partial_{\lambda}\tilde{S}$ and
$-\dot{\hat{\xi}} = -K_{\lambda}$ as required)
\\[2mm]\indent Now in our semiclassical action principle of [3;4],
based on the Jevicki-Yoneya action principle of [18;49], the quantity
$S - \xi$ was introduced via heuristic considerations and served very
well (see below).  It's origin was the use of $log(W)\sim -H$ as an
action ingredient in [18;49], corresponding to $W\sim exp(-\frac{1}
{\epsilon}H)$ in the semiclassical version.  Thus a basic action was
posited to be $Sp(H)\sim \int \oint H\frac{dk}{2i\pi}$ in [18;49] and
we observed that $Wexp(\xi) = w\sim exp(\frac{1}{\epsilon}S)$ corresponds
to $-H = S - \xi$.  This identification has an interesting interpretation
in terms of the quasiclassical limit of the Sato equation
\medskip
$$\leqno(5.6)\quad (\partial_{n}W)W^{-1} = -L^{n}_{-}$$
\medskip
\noindent Thus from (4.7)-(4.9), writing e.g. $W(T,\epsilon) = 1 +
\sum^\infty_{1} w_{n}(\epsilon,T)(\epsilon\partial)^{-n},$ one obtains
\medskip
$$\leqno(5.7)\quad \epsilon(\partial_{n}W)W^{-1}\to -\lambda^{n}_{-}$$
\medskip
\noindent Note formally for $W\sim exp(-\frac{1}{\epsilon}H), \epsilon
\partial_{n}log(W) = \epsilon(\partial_{n}W)W^{-1} = -\partial_{n}H\sim
\partial_{n}(S - \xi)$ while directly from (4.7) or (5.1)-(5.3)
\medskip
$$\leqno(5.8)\quad \partial_{n}(S - \xi) = {\cal B}_{n} - \lambda^{n}
= \lambda^{n}_{+} - \lambda^{n} = -\lambda^{n}_{-}$$
\medskip
\noindent Consequently one has (cf. also [42;43] for some related ideas)
\\[3mm]\indent {\bf THEOREM 5.1.} $\,$ The term $S - \xi$ arising in the
semiclassical action principle of [3;4] can be connected to W via a
semiclassical limit of the Sato equation and has the form $\partial_{n}
(S - \xi) = -\lambda^{n}_{-}.$
\\[3mm]\indent It would also be of interest to investigate the sense
in which $S - \xi$ relates to action (via residue calculation).
\\[3mm]\indent {\bf REMARK 5.2.} $\,$ It is clear that the Maslov
canonical operator (cf. [28]) is connected with semiclassical soliton
theory and we wrote this out in an earlier version of this paper.
There does seem to be good motivation for pursuing this (cf. also [43]).
\\[3mm]\indent {\bf REMARK 5.3}.$\,\,$ We mention a few connections
of our work to the development in [43], which has just come to our
attention.  We hope to return to this in more detail at another time.
We note that in [43] one writes $\lambda = exp(ad(\phi))P,\,\,
(P\sim k,\lambda\sim {\cal L})$, via a dressing function $\phi$,
with $exp(ad(\phi))X = X + \sum^{\infty}_{1}
V_{i+1}\lambda^{-i-1}$.  This last expression corresponds to our
relation $WxW^{-1} = G\to \hat {\xi} = X + \sum^\infty_{1}V_{i+1}
\lambda^{-i-1}$, derived in [3;4] (cf. also (4.11)).  Thus $\lambda =
exp(ad(\phi))P$ and $\hat {\xi} = exp(ad(\phi))X$ where $(\lambda,-\hat{\xi})$
are the action-angle variables of [3;4;24], and $ad\phi(\psi) = \{\phi,
\psi\}$.  There is also a connection to [49] (cf. also [3;4]) where
$\lambda = lim \, exp(-\frac{1}{\epsilon}H\circ)Pexp(\circ
\frac{1}{\epsilon}H)$
is used with $(f\circ g)(x,k) = f(x,k)exp(\epsilon\overleftarrow
{\partial}_{k}\overrightarrow {\partial}_{x})g(x,k)$ and $\{f,g\} =
\frac{1}{\epsilon}(f\circ g - g\circ f) (W_{\epsilon}\sim
exp(-\frac{1}{\epsilon}H)$).  Finally let us remark that in [43] one
provides an important quasiclassical limit of certain Hirota equations
via a quasiclassical differential Fay identity which has as a consequence
the formula (a minus sign typo is corrected)
\medskip
$$\leqno(5.9)\quad \sum^\infty_{m,n = 1}\mu^{-m}\lambda^{-n}\frac
{\partial_{n}\partial_{m}F}{mn} = log[1 + \sum^\infty_{1}
\frac{\lambda^{-n} - \mu^{-n}}{\mu - \lambda}\frac{\partial\partial_{n}F}
{n}]$$
\medskip\noindent where $log(\tau)^{dKP}\sim F$ and $\tau(\epsilon,T)\sim
exp(\frac{1}{\epsilon^{2}}F)$ (cf. (4.12).
This formula has an interesting version in terms of
the quantity $P(\mu) - P(\lambda)$ which plays an important role in the
Hamilton-Jacobi theory of [24] (cf. also [3;4]).  Thus in (4.6) one
writes $P(\lambda) = \lambda - \sum^\infty_{1}P_{n}\lambda^{-n}$
where (cf. (4.8)) $-P_{n} =
\partial S_{n+1}$ and $-nS_{n+1} = \partial_{n}F$
which implies $\partial\partial_{n}F = -n\partial S_{n+1} = nP_{n}$
(note also $-P_{n} = q_{n+1}$ in the notation of [42;43]).  Hence
in (5.9) $\sum^\infty_{1}\lambda^{-n}\frac{\partial\partial_{n}F}
{n} = \sum^\infty_{1}\lambda^{-n}P_{n} = \lambda - P(\lambda)$
which implies
\medskip
$$\leqno(5.10)\quad \sum^\infty_{m,n=1}\mu^{-m}\lambda^{-n}
\frac{\partial_{n}\partial_{m}F}{mn} = log[\frac{P(\mu) - P(\lambda)}
{\mu - \lambda}]$$
\medskip\noindent We anticipate that this formula might prove interesting
in terms of phase space geometry, topological field theory, etc. (cf.
[2-4;13;24;25;42;43;49]).  In particular, in [6] we show how to
extract the dispersionless Hirota type equations from (5.16), using
(5.17).  These are nonlinear partial differential equations involving
$\partial_n\partial_m F,$ and $\partial\partial_n F$ which should
characterize F.

\noindent{\bf APPENDIX A. Coherent states.}
\\[5mm]\indent We collect here a few remarks and formulas concerning
coherent states and various representations in quantum mechanics (cf.
[25;41;49] for background material).  We think of $Q\sim \hat{q}_{h}, P\sim
\hat{p}_{h}$ as in (2.6) with $a\sim a_{h}, a^{\dagger}\sim
a^{\dagger}_{h}.$  There are various representations of vectors in
terms of coordinates, momenta, number operators, coherent states, etc.
and we describe this briefly.  Then, given boson operators $a,a^{\dagger}$
with $[a,a^{\dagger}] = 1$ one can choose a vacuum vector $|0>$ with
$a|0> = 0$ and normalized vectors $(<0|0> = 1)$
\medskip
$$\leqno(A.1)\quad |n> = (a^{\dagger})^{n}|0>/\sqrt{n!};\,
<m|n> = \delta_{m,n}$$
\medskip
\noindent As an example, for $z\in S^{1}, a\sim \partial_{z},
a^{\dagger}\sim z, |n> = \frac{z^{n}}{\sqrt{n!}}, <n|m> = \frac{1}{2i\pi}
\int z^{n}\bar{z}^{m}(\frac{dz}{z}).$ The number operator is $a^{\dagger}a$
with $a^{\dagger}a|n> = n|n>, a|n> = \sqrt{n}|n-1>,$ and $a^{\dagger}|n>
= \sqrt{n+1}|n+1>.$  In a more quantum mechanical spirit (cf. [44]) one
has position and momentum representations via (recall $\hat{Q}$ means operator
Q)
$\hat{Q}|Q'> = Q'|Q'>, <Q''|Q'> = \delta(Q'' - Q'), 1 = \int dQ'(|Q'><Q'|),
(\int = \int^{\infty}_{-\infty}),$ and for states $|\alpha>, |\beta>,
|\alpha> = \int dQ'(|Q'><Q'|\alpha>),$ with
\medskip
$$\leqno(A.2)\quad <Q'|\alpha> = \psi_{\alpha}(Q'); <\beta|\alpha> =$$
$$\int dQ'<\beta|Q'><Q'|\alpha> = \int dQ'\bar{\psi}_{\beta}(Q')
\psi_{\alpha}(Q')$$
\medskip
\noindent Also in general $<\beta|\hat{A}|\alpha> = \int \int dQ'dQ''
\bar{\psi}_{\beta}<Q'|\hat{A}|Q''>\psi_{\alpha}(Q'').$  For the momentum
representation $P\sim\hat{p}_{h}$ with $i\hat{P} = h\partial_{Q}),
\hat{P}|P'> = P'|P'>,\\ <P'|P''> = \delta(P' - P''), 1 = \int dP'
(|P'><P'|), |\alpha> = \int dP' (P'><P'|\alpha>,$\\ and
\medskip
$$\leqno(A.3)\quad <P'|\alpha> = \phi_{\alpha}(P'); <Q'|\hat{P}|\alpha> =
-ih\partial_{Q'}<Q'|\alpha>;$$
$$<Q'|\hat{P}|Q''> = -ih\partial_{Q'}\delta(Q' - Q''); <Q'|P'> =
\frac{1}{\sqrt{2h\pi}}e^{\frac{i}{h}P'Q'}$$
\medskip
\noindent In this context vacuum vectors $|0>$ such that $a|0> = 0$
 ($\sim a_{h}|0> = 0$) can be represented in a peaked state or
Schr$\ddot o$dinger form via the coordinate Q.  Thus (cf. [36]) from
$<Q'|a|0> = 0$ we have $(Q' + h\partial_{Q'})<Q'|0> = 0$ or
\medskip
$$\leqno(A.4)\quad <Q'|0> = (h\pi)^{-\frac{1}{4}}e^{-\frac{1}{2h}
Q'^{2}}; <Q'|n> =$$
$$\frac{(Q' - h\partial_{Q'})^{n}}{(h\pi)^{\frac{1}{4}}
\sqrt{n!(2h)^{n}}}e^{-\frac{1}{2h}Q'^{2}} = e^{-\frac{1}{2h}Q'^{2}}
H_{n}(\frac{Q'}{\sqrt{h}})/(h\pi)^{\frac{1}{4}}\sqrt{n!2^{n}}$$
\medskip
\noindent These will be referred to as oscillator eigenfunctions or the
Schr$\ddot o$dinger representation ($H_{n}\sim$ Hermite polynomial).
\\[5mm]\indent Next, coherent states are defined via
\medskip
$$\leqno(A.5)\quad |z> = D(z)|0> = e^{-\frac{1}{2}|z|^{2}}\sum^\infty_{0}
\frac{z^{n}}{\sqrt{n!}}|n>;
D(z) = e^{za^{\dagger}_{h} - \bar{z}a_{h}}; z = \frac{1}{\sqrt{2h}}(q + ip)$$
\medskip
\noindent We write also $|z>\sim |p,q>$ and set
\medskip
$$\leqno(A.6)\quad U(p,q) = e^{\frac{i}{h}(p\hat{q}_{h} -q\hat{p}_{h})}\sim
e^{\frac{i}{h}(p\hat{Q} - q\hat{P})}$$
\medskip
\noindent Note $a_{h} = \frac{1}{\sqrt{2h}}(\hat{q}_{h} + i\hat{p}_{h}),
a^{\dagger}_{h} = \frac{1}{\sqrt{2h}}(\hat{q}_{h} - i\hat{p}_{h}),$
so $za^{\dagger}_{h} - \bar{z}a_{h} = (\frac{i}{h})(p\hat{q}_{h} -
q\hat{p}_{h}).$  Thus $|z>\sim U(p,q)|0>$ and one records
\medskip
$$\leqno(A.7)\quad a_{h}|z> = z|z>; D^{\dagger}(z) = D^{-1}(z) = D(-z);
D^{\dagger}(z)a_{h}D(z) = a_{h} + z;$$
$$D(z) = e^{-\frac{1}{2}|z|^{2}}e^{za^{\dagger}_{h}}e^{-\bar{z}a_{h}} =
e^{\frac{1}{2}|z|^{2}}e^{-\bar{z}a_{h}}e^{za^{\dagger}_{h}}; <z|z'> =
e^{-\frac{1}{2}|z|^{2} + \bar{z}z' -\frac{1}{2}|z'|^{2}};$$
$$<z'|a_{h}|z> = z<z'|z>; <z'|a^{\dagger}_{h}|z> = \overline {<z|a_{h}|z'>} =
\bar{z}<z'|z>;$$
$$1 = \int (|z><z|)d\mu (d\mu = \frac{1}{\pi}dz_{1}dz_{2}); D(z)|\zeta>
= e^{iIm(z\bar{\zeta})}|z+\zeta>$$
\medskip
\noindent The measure $d\mu$ will change for $z\in S^{1}$ or for other
constraints. (cf. (3.11)-(3.14)).  One is also interested in the Bargman
representation through $\tilde{\psi}(z)$ where
\medskip
$$\leqno(A.8)\quad \psi(z) = <z|\psi> = e^{-\frac{1}{2}|z|^{2}}
\sum^\infty_{0}\frac{\bar{z}^{n}}{\sqrt{n!}}<n|\psi> = e^{-\frac{1}{2}
|z|^{2}}\tilde{\psi}(\bar{z})$$
\medskip
\noindent Here $\tilde{\psi}(\bar{z})$ is an analytic function of $\bar{z}$.
Finally
let us record a formula from [40] (for $[a,a^{\dagger}] = 1)$
\medskip
$$\leqno(A.9)\quad [a,g(a^{\dagger})] =
(\frac{\partial}{\partial a^{\dagger}})g(a^{\dagger}); [f(a),a^{\dagger}] =
(\frac{\partial}
{\partial a})f(a)$$
\\[3mm]\noindent {\bf APPENDIX B. Operator structure.}
\\[.5cm]\indent We elaborate here on (4.35)-(4.37) and Corollary 4.4.
First observe from [47] that given $\hat {\Lambda}\sim \zeta_{0}$
(cf. remarks before (2.16)) and $\zeta = Ad^{*}_{u}\zeta_{0}$ defined
as in (2.18), the coordinates of $\zeta$ can be taken as $\zeta_{1} =
\zeta^{0}_{1} + p\zeta_{3}, \zeta_{2} = \zeta^{0}_{2} + q\zeta_{3},$
and $\zeta_{3} = \zeta^{0}_{3}$ (with $\zeta^{0}_{3} = 1$ say).
Hence (p,q) (note the order) serve as coordinates on a given coadjoint
orbit $\Gamma = \{Ad^{*}_{u}\zeta_{0}\}$ and if we think of $U\sim
U(\pi,\xi)\sim D(z)$ with $z = \frac{1}{\sqrt{2h}}(\xi + i\pi)$ as
in (4.23), then $(\pi,\xi)$ determines coordinates on $\Gamma\subset
\tilde {g}^{*}$ (cf. (2.15) - $\tilde {g}\sim \tilde {g}_{1}$ or a generic
$\tilde {g}_{h}$).  Thus $U\in G_{h}$ gives rise to
$\zeta\in \Gamma\subset  \tilde {g}^{*}$ and $\Gamma\sim \frac{G_{h}}
{H_{0}}$ ($H_{0} = H_{\zeta_{0}}$) as indicated before (2.15).  The
map $J: [U]\to \zeta: \frac{G_{h}}{H_{0}} = M\to \Gamma$ is in fact
a sort of momentum map (cf. [8]).  Recall one defines J via
$J: M\to \tilde {g}^{*}; J_{*}: TM\to T\tilde {g}^{*}\simeq  \tilde {g}^{*};
J^{*}: T^{*}\tilde {g}^{*}\simeq \tilde {g}\to T^{*}M; J^{*}\xi
= d\hat {J}(\xi)$
for $\xi\in \tilde {g}$ where $\hat {J}: \tilde {g}\to C^{\infty}(M)$
and $\xi_{M}(m) = X_{\hat {J}(\xi)}(m)$ for $\xi_{M}(m) = D_{t}
\phi(exp(t\xi))m|_{t=0}$ and $X_{f}(g) = \{g,f\}$.  In order to erect
some heuristic structure we will not belabor details here.  Now consider
operators $\hat {A}, \hat {A}_{QC}$ as in (4.35)-(4.36) so that
\medskip
$$\leqno(B.1)\quad <z|\hat {A}|z> = \int \int d\tilde {p}d\tilde {q}
\tilde {f}(\tilde {p},\tilde {q})<z|e^{i(\tilde {p}\hat {Q} - \tilde {q}
\hat {P})}|z>;$$
$$<z|\hat {A}_{QC}|z> = \int \int d\tilde {p}d\tilde {q}
\tilde {f}(\tilde {p},\tilde {q})<z|e^{(\frac{i}{\epsilon})(\tilde {p}
\hat {Q} - \tilde {q}\hat {P})}|z>$$
\medskip
\noindent From Theorem 4.3 we know e.g. that $<z|exp(\frac{i}{\epsilon})
(\tilde {p}\hat {Q} - \tilde {q}\hat {P})|z>\sim e^{-\frac{1}{2}}
exp[(\frac{i}{\epsilon})(\tilde {p}\xi - \tilde {q}\pi)]$ ($z\sim
(\xi,\pi)$) and we assume the error term is suitably small etc. when
writing (B.1) (for $\hat {A}$ with $<z|exp[i(\tilde {p}\hat {Q} -
\tilde {q}\hat {P})]|z>$ this is surely OK for reasonable $\tilde {f}$).
Now from remarks before (2.16) we can write (here think of $\hat {\Lambda} =
\frac{\Lambda}{\epsilon^{2}}\sim  \frac{\lambda}{\epsilon^{2}}\in
\tilde {g}_{h}, \lambda\sim (-\tilde {q},\tilde {p})$ and the $e_{3}$
terms can be neglected as in sections 2 and 4)
\medskip
$$\leqno(B.2)\quad lim (\frac{\epsilon^{2}}{i})<z|\hat {\Lambda}|z> =
<Ad^{*}_{U(\pi,\xi)}\zeta_{0},\lambda>$$
$$= <\zeta,\lambda>\sim <(\pi,\xi),(-\tilde {q},\tilde {p})> =
\tilde {p}\xi - \tilde {q}\pi$$
\medskip
\noindent ($\zeta\sim (\pi,\xi)$ comes from z, $\zeta_{0}\sim (0,0),$ and
$\lambda\sim (-\tilde {q},\tilde {p})$).  Consequently our quasiclassical
operator $\hat {A}_{QC}$ has the approximate symbol
\medskip
$$\leqno(B.3)\quad <z|\hat {A}_{QC}|z>\sim \int \int d\tilde {p}d\tilde {q}
\tilde {f}(\tilde {p},\tilde {q})e^{\frac{i}{\epsilon}<\zeta,\lambda>} =
a(\frac{\zeta}{\epsilon})\sim $$
$$\int_{\tilde {g}} d\lambda \tilde {F}(\lambda)
e^{\frac{i}{\epsilon}<\zeta,\lambda>}; <z|\hat {A}|z>\sim
\int_{\tilde {g}} d\lambda\tilde {F}(\lambda)e^{i<\zeta,\lambda>} = a(\zeta)$$
\medskip
\noindent where $\tilde {g}\sim \tilde {g}_{1}$ say, $\zeta\in \Gamma,$
and $e^{-\frac{1}{2}}$ terms are ignored for convenience.  One can
in fact take equivalence classes $[\lambda]\in \tilde {g}$,
as indicated before (2.15) so that the integrals in (B.3) are over
$T^{*}_{\zeta}\Gamma$ but $\tilde {g}$ should be retained for inversion.
Thus one enters the realm of coherent state transforms, oscillatory
integrals, Weyl-Wigner-Moyal theory, etc. with many possibilities
for further development (cf. [14;28;29;39]).  In particular
one can recover $\tilde {F}(\lambda)$ via Fourier inversion which
amounts to determining the asymptotic forms for $\hat {A}$ or $\hat {A}_{QC}$
from their diagonal symbols.

\begin{thebibliography}{77}
\bibitem{am92}
M. Adler and P. vanMoerbeke,
Comm. Math. Physics, 147 (1992),25-56
%
\bibitem{ak}
S. Aoyama and Y. Kodama,
Phys. Lett. B, 278 (1992), 56-62; 295 (1992), 190-198; hep-th 9404011
%
\bibitem{rc}
R. Carroll,
Remarks on dispersionless KP, KdV, and 2-D gravity, Jour. Nonlinear Science,
to appear
%
\bibitem{rc2}
R. Carroll,
On canonical variables in soliton hierarchies, Teor. i Matem. Fizika, to
appear
%
\bibitem{uv}
R. Carroll,
Mathematical Physics, North-Holland, 1988
%
\bibitem{hv1}
R. Carroll,
On dispersionless Hirota type equations, Proc. NEEDS '94, World
Scientific, to appear (hep-th 9410063)
%
\bibitem{rc6}
R. Carroll,
Applicable Anal., 49 (1993), 1-31; Symmetries, Sato theory, and tau
functions, Applicable Anal., to appear
%
\bibitem{rc7}
R. Carroll,
Topics in soliton theory, North-Holland, l991
%
\bibitem{djmk}
E. Date, M. Jimbo, M. Kashiwara, and T. Miwa,
Proc. RIMS Sympos., World Scientific, 1983, pp. 39-119
%
\bibitem{de}
S. deBievre and J. Gonzalez,
Quantization and coherent states, World Scientific, 1993, pp. 152-157
%
\bibitem{dik}
L. Dickey,
Soliton equations and Hamiltonian systems, World Scientific, 1991
%
\bibitem{qd1}
L. Dickey,
Mod. Phys. Lett. A, 8 (1993), 1259-1272; 1357-1377
%
\bibitem{du}
B. Dubrovin,
Verdier Memor. Conf., Birkhauser, 1993, pp. 313-359; Nuc. Phys. B, 379
(1992), 627-689; Comm. Math. Phys., 145 (1992), 195-207
%
\bibitem{xc1}
G. Folland,
Harmonic analysis in phase space, Princeton, 1989
%
\bibitem{gr}
P. Grinevich and A. Orlov,
Problems in modern quantum field theory, Springer, 1989, pp. 86-106
%
\bibitem{ha}
G. Hagedorn,
Comm. Math. Phys., 71 (1980), 77-93
%
\bibitem{hep}
K. Hepp,
Comm. Math. Phys., 35 (974), pp. 265-277
%
\bibitem{jy}
A. Jevicki and T. Yoneya, Mod. Phys. Lett. A, 5 (1990), 1615-1621
%
\bibitem{ji}
M. Jimbo and T. Miwa,
Publ. RIMS, 19 (1983), 943-1001
%
\bibitem{kr}
V. Kac and A. Raina,
Highest weight representations of infinite dimensional Lie algebras,
World Scientific, 1987
%
\bibitem{ka}
A. Kaneko,
Introduction to hyperfunctions, Kluwer, 1988
%
\bibitem{kl}
J. Klauder and E. Sudarshan,
Fundamentals of quantum optics, Benjamin, 1968
%
\bibitem{ko}
Y. Kodama,
Prog. Theor. Phys., Supp. 94 (1988), 184-194; Phys. Lett. A, 129 (1988),
223-226
%
\bibitem{kog}
Y. Kodama and J. Gibbons,
Proc. Inter. Workshop Nonlinear and Turbulent Processes in Physics,
Kiev, 1989, pp. 166-180
%
\bibitem{kri}
I. Krichever,
Comm. Math. Phys., 143 (1992), 415-429;
Comm. Pure Appl. Math., 47 (1994), 437-475
%
\bibitem{krs}
H. Kuratsugi, A. Inomata, and C. Gerry,
Path integrals and coherent states of SU(2) and SU(1,1), World Scientific,
1992
%
\bibitem{lx}
P. Lax, C. Levermore, and S. Venakidis,
Important developments in soliton theory, Springer, 1993, pp. 205-241
%
\bibitem{msl}
V. Maslov and M. Fedoryuk,
Semiclassical approximation in quantum mechanics, Reidel, 1981
%
\bibitem{rm}
E. Meinrenken,
Jour. Phys. A, 27 (1994), 3257-3265;
Rep. Math. Phys., 31 (1992), 279-295
%
\bibitem{mu}
M. Mulase,
Jour. Diff. Geom., 19 (1984), 403-430
%
\bibitem{nu}
D. Mumford,
Tata Lectures on Theta, II, Birkhauser, 1984
%
\bibitem{or}
A. Orlov and E. Schulman,
Lett. Math. Phys., 12 (1986), 171-179; Teor. i Matem. Fizika,
64 (1985), 323-328
%
\bibitem{pr}
A. Perelomov,
Generalized coherent states and their applications, Springer, 1986
%
\bibitem{rs}
M. Rasetti and G.D'Ariano, Integ. systems in statistical mechanics,
World Scientific, 1985, pp. 143-152
%
\bibitem{rsm}
M. Rasetti, G. D'Ariano, and A. Montorsi, Nuovo Cimento, 11 (1989), 19-28
%
\bibitem{sak}
J. Sakurai,
Modern quantum mechanics, Benjamin, 1985
%
\bibitem{sh}
L. Schulman,
Techniques and applications of path integrals, Wiley, 1981
%
\bibitem{sh}
T. Shiota,
Invent. Math., 83 (1986), 333-382
%
\bibitem{sch}
M. Shubin and F. Berezin,
The Schrodinger equation, Kluwer, 1991
%
\bibitem{st}
W. Steeb and K. Kowalski,
Nonlinear dynamical systems and Carleman linearization, World Scientific, 1991
%
\bibitem{sw}
M. Swanson,
Path integrals and quantum processes, Academic Press, 1992
%
\bibitem{TT}
K. Takasaki and T. Takebe,
Inter. Jour. Mod. Phys. A, Supp. 1992, pp. 889-922
%
\bibitem{tt3}
K. Takasaki and T. Takebe,
hep-th 9405096
%
\bibitem{tt}
K. Takasaki,
Jour. Geom. Phys., 14 (1994), 111-120
%
\bibitem{th}
W. Thirring,
Quantum mechanics of atoms and molecules, Springer, 1981
%
\bibitem{ts}
A. Tsuchiya, N. Kawamoto, Y. Namikawa, and Y. Yamada,
Comm. Math. Phys., 116 (1988), 247-308
%
\bibitem{ya}
L. Yaffe,
Rev. Mod. Phys., 54 (1982), 407-435
%
\bibitem{ya2}
L. Yaffe and F. Brown,
Nuc. Phys. B, 271 (1986), 267-332
%
\bibitem{yo}
T. Yoneya,
Comm. Math. Phys., 144 (1992), 623-638; Inter. Jour. Mod. Phys. A, 7 (1992),
4015-4038
%
\bibitem{za}
W. Zhang and D. Feng,
Coherent states: Past, Present, and Future, World Scientific, 1994, pp.
561-580
%
\end{thebibliography}
\end{document}